\documentclass[showpacs,aps,twocolumn]{revtex4-1}
\usepackage{amsmath}
\usepackage{amstext}
\usepackage{amsopn}
\usepackage{amsfonts}
\usepackage{bbm}
\usepackage{accents}
\usepackage{empheq}
\usepackage{graphicx}
\usepackage{epsf}
\usepackage{graphics}
\usepackage{cancel}[]
\usepackage{amssymb}
\usepackage{subfigure}
\usepackage{color}

\begin{document}
\title{The response of laser interferometers to a gravitational wave}

\author{ Adrian Melissinos$^{a}$ and Ashok Das$^{a,b}$}
\affiliation{$^{a}$ Department of Physics and Astronomy, University of Rochester, Rochester, NY
14627-0171, USA}
\affiliation{$^b$ Saha Institute of Nuclear Physics, 1/AF Bidhannagar, Calcutta 700064, India}


\begin{abstract}
Laser interferometer detectors are now widely used in an attempt to
detect gravitational waves (gw). The interaction of the gw with the
light circulating in the interferometer is usually explained in
terms of the motion of the ``free" mirrors that form the
interferometer arms. It is however instructive to show that the same
result can be obtained by simply calculating the propagation of an
electromagnetic plane wave between ``free mirrors" in the curved
space-time induced by the gw. One finds that the plane wave acquires
frequency modulation sidebands at the gw frequency, as would be
expected from the absorption and emission of gravitons from and to
the gw. Such sidebands are completely equivalent to the
time-dependent phase shift imposed on the plane wave, that follows
from the conventional calculation.
\end{abstract}

\maketitle

\section{Introduction}

In discussing the operation of gravitational wave (gw)
interferometric detectors \cite{Abramovici, Forward} it is usually
stated that the distance between the beam splitter and the mirrors
at the end of the arms is changed by the gw
\cite{Bluebook,Saulsonbook}. When light beams propagate in the arms,
the change in the proper length of the two arms results in a phase
shift between the light beams that return to the beam splitter. The
phase shift is the physical observable that indicates the presence
of a gw. The above statement is valid if the mirrors are free to
move (along the axis of the arms) and it is expressed in the
laboratory frame of reference, what is referred to as the Local
Lorentz (LL) gauge. It is however much easier to calculate the
resulting phase shift in the Transverse Traceless (TT) gauge. In the
TT gauge the gw has a very simple form and the coordinates of free
particles (i.e. of the mirrors) are not changed in the presence of
the gw, even though their separation does change
\cite{AJP1,AJPSaulson}.

 The geometry
of space is defined by the infinitesimal interval \cite{MTW}

\begin{equation}
ds^2 = g_{\mu\nu} dx^{\mu} dx^{\nu}, \qquad \mu, \nu, =
0,1,2,3,
\end{equation}

\noindent where summation over repeated indices is implied. For weak
fields we write the metric tensor as

\begin{equation}
g_{\mu\nu} = \eta_{\mu\nu} + h_{\mu\nu}, \qquad
h_{\mu\nu} \ll 1.
\end{equation}

\noindent The flat-space (Minkowski) metric $\eta_{\mu\nu}$ is
chosen to be
\begin{equation}
\eta_{\mu\nu} = \begin{pmatrix}  -1 &0  & 0 & 0\\
                     0 &1 &0  &0  \\
                     0 &0  &1 &0  \\
                     0 &0  &0  &1 \\
\end{pmatrix}.
\end{equation}

Consider the interferometer in the $x-y$ plane and that a
gravitational wave of angular frequency $\Omega$ is incident along
the $z$-axis. In the TT gauge the potential $h_{\mu\nu}$ is given
by the real part of the following expression

\begin{equation}
h_{\mu\nu} = e^{-i(\Omega t + k_{\Omega}z)} \begin{pmatrix}  0 & 0  & 0 & 0\\
                     0 & h_{+} &h_{\times}  &0  \\
                     0 & h_{\times}  &-h_{+} &0  \\
                     0 & 0  &0  &0 \\
\end{pmatrix}.
\end{equation}

\noindent  $h_{+}$ and $h_{\times}$ are the (real) amplitudes of the
``parallel" and ``cross" polarization states of the gw. To be
specific we will place the origin of the coordinates at the beam
splitter and orient the interferometer arms along the $x$- and $y$-
axes, as shown in Fig. (1). In the next section we calculate, in the
TT gauge, the round trip time for the propagation of light from the
beam splitter to the end mirror and back to the origin, in the
presence of a gw. We will then obtain the same result by
considering, again in the TT gauge, the direct interaction of the gw
with the light circulating in the arms.

\begin{figure} [h!]
\centering
\includegraphics[width=70mm,height=70mm]{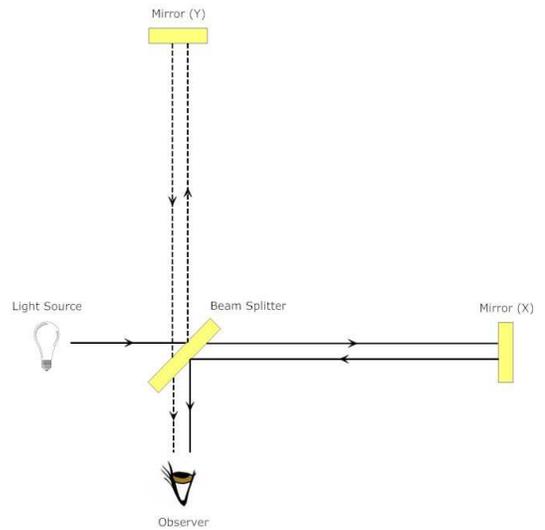}
\caption{ Configuration of a simple Michelson Interferometer.}
\end{figure}

\section{Conventional calculation}

We will work in the TT gauge so the coordinates of the mirrors are
unchanged by the presence of the gravitational wave. We will
calculate the time it takes for light leaving the beam splitter (the
origin) to reach the end mirror, where it is reflected, and to
return to the the beam origin. For light $ds^2 = 0$, and since the
propagation is in the $x-y$ plane, Eq. (1) reduces to

\begin{equation}
g_{00} c^2 dt^2 + g_{11} dx^2 + g_{22} dy^2 + 2 g_{12} dx dy =
0.\nonumber
\end{equation}

\noindent For propagation only along the $x$-axis, and since we can
set $z = 0$,

\begin{align}
c\left |\frac{dt}{dx}\right | = \left |\frac{g_{11}}{g_{00}}\right
|^{1/2} = (1 + h_{11}(t))^{1/2} \simeq  1 + \frac{1}{2}h_{+}
e^{-i\Omega t} .
\end{align}

\noindent The time interval for a round trip is obtained by
integrating from the origin to the end of the arm, $x = L$, and back
to the origin. If the light leaves at $t = t_0$ and returns at $t =
t_r$

\begin{align}
t_r = t_0 + &\int\limits_0^L \left|\frac{dt}{dx}\right|dx -
\int\limits_L^0 \left|\frac{dt}{dx}\right|dx,
\end{align}

\noindent where the minus sign before the second integral accounts for the
fact that on the return trip $dt/dx = - \left|dt/dx\right|$. In
carrying out the integration we can replace $t$ in the exponential
by $(x/c)$, because any corrections will be of second order in $h$.
We also  define the one way travel time in the absence of the gw as
$T = L/c$. Thus

\begin{align}
 t_r &= t_0 + \frac{1}{c}\int\limits_0^L
\left[1 + \frac{h_{+}}{2}e^{-i\Omega(t_0 + x/c)}\right]dx
 \nonumber \\ & \qquad \ - \frac{1}{c}\int\limits_L^0 \left[1 +
\frac{h_{+}}{2}e^{-i\Omega(t_0 + 2T - x/c)}\right]dx \nonumber \\
 &= t_0 + \frac{2L}{c} + \frac{h_{+}}{2i\Omega}
e^{-i\Omega t_0} \left[ 1 - e^{-i2\Omega T}\right]\nonumber\\
&= t_0 + \frac{2L}{c} + h_{+}T e^{-i\Omega (t_0 + T)}\frac{\rm {sin}
\Omega T }{\Omega T}.
\end{align}

\noindent This suggests that we can write

\begin{equation}
h_{11}(t_r) = h_{+}e^{-i\Omega t_r} \simeq h_{+} e^{-i\Omega(t_0 +
2T)}, \nonumber
\end{equation}

\noindent where we have neglected higher order terms in the metric.
As a result, the travel time along the $x$-axis can be expressed as

\begin{equation}
\Delta t_x(t_r) = t_r - t_0 = \frac{2L}{c} + h_{11}(t_r) \frac{L}{c}
\frac{\rm{sin}(\Omega T)}{\Omega T} e^{i\Omega T}.
\end{equation}

\noindent For the $y$-axis we obtain the same result with
$h_{11}(t)$ replaced by $h_{22}(t) = - h_{11}(t)$. Thus the
difference in travel time between the two arms is

\begin{equation}
\Delta t(t) = \Delta t_x(t) - \Delta t_y(t) = 2 [h_{+}e^{-i\Omega
t}] \frac{L}{c} \frac{\rm{sin}(\Omega T)}{\Omega T} e^{i\Omega T}.
\end{equation}\\

A difference in the time of arrival of the light from the two arms,
implies a phase shift between the two fields. If the angular
frequency of the carrier is $\omega_0$, the phase shift
corresponding to a delay $\Delta t$, is $\Delta \phi = \omega_0
\Delta t = c k_0 \Delta t$, where $k_0 = \omega /c$ is the
wavenumber of the carrier. Therefore when the gw is normally
incident on the interferometer plane and when the polarization of
the gw is along the interferometer axes, the observable phase shift
is

\begin{equation}
\Delta \phi(t) = 2 k_0 L h_{+} \frac{\rm{sin}(\Omega T)}{\Omega T}
e^{-i\Omega(t -  T)},
\end{equation}

\noindent where following the convention  adopted for Eq. (4),
$\Delta \phi (t)$ is given by the real part of Eq. (10). For a more
detailed discussion of this derivation, see \cite{Malik1}.

\section{The field equations in the presence of the gw}

We will now show that the same result can be obtained by considering
the direct interaction of the gravitational wave with the light
propagating in the interferometer arms. In this section we find the
equations for the electromagnetic field in the presence of the gw,
and in the next section we solve the equations for light propagating
in the $x$ and/or $y$ arms. We find the presence of (frequency
modulation) sidebands on the carrier displaced by the gw frequency
$\Omega $. This is equivalent to the phase shift found in Eq. (10).

In flat space Maxwell's inhomogeneous equations can be written in
the manifest covariant form as

\begin{equation}
\partial_{\mu} F^{\mu \nu} = j^{\nu},\label{mweqn}
\end{equation}
where the field strength tensor is defined as
\begin{equation}
F_{\mu\nu} = \partial_{\mu}A_{\nu} - \partial_{\nu}A_{\mu}.\label{fieldstrength}
\end{equation}
In a curved space  and in the absence of sources, Eq. \eqref{mweqn} is replaced
by \cite{Landau_Lifshitz}

\begin{equation}
\partial_{\mu}\left(\sqrt{-g}\ g^{\mu \lambda} g^{\nu \rho} F_{\lambda \rho}\right)
= 0,\label{mwcurvedeqn}
\end{equation}
where $g = {\rm{det}}(g_{\mu \nu})$ with the field strength tensor still defined
by Eq. \eqref{fieldstrength}. The dynamical equation \eqref{mwcurvedeqn} can be
derived as the Euler-Lagrange equation from the action described by the Lagrangian density

\begin{align}
{\cal L} & = -\frac{1}{4}\sqrt{-g}\ g^{\mu \lambda} g^{\nu
\rho}F_{\mu \nu} F_{\lambda \rho}
\nonumber\\
& =-\frac{1}{4}\sqrt{-g}\ g^{\mu \lambda} g^{\nu \rho}
\left(\partial_{\mu} A_{\nu} -
\partial_{\nu} A_{\mu}\right)\left(\partial_{\lambda} A_{\rho} - \partial_{\rho}
A_{\lambda}\right).
\end{align}

We will again work in the TT gauge and take the gw to be the same as before.
In the TT gauge, in the weak-field approximation $\sqrt{-g} = 1$, and keeping
only terms linear in $h_{\mu \nu}$,
Eq. \eqref{mwcurvedeqn} reads

\begin{equation}
\partial_{\mu} F^{\mu \nu} - \partial_{\mu}(h^{\mu \lambda} F_{\lambda}^{.\ \nu})
- \partial_{\mu}(h^{\nu \rho} F^{\mu}_{.\ \rho}) = 0.
\end{equation}

\noindent The first term is the same as Eq. \eqref{mweqn} with
$j^{\nu} = 0$, and describes the propagation of the free
electromagnetic field. The next two terms act as sources that give
rise to new electromagnetic fields generated by the interaction of
the gw with the field $F^{\mu \nu}$ of the light circulating in the
arms. For the choice of gw (normal incidence toward the negative
z-axis), expansion of Eq. (14) in $\mathbf{E}\ {\rm{and}}\ \mathbf
{B}$ field components yields the four equations

\begin{align}
\nu = 0:\quad & \mbox{\boldmath$\nabla$}  \cdot \mathbf {E} - h_{+}e^{-i(\Omega t + k_{\Omega}z)}(\partial_{x} E_{x} - \partial_{y} E_{y})\nonumber\\
&\quad  -  h_{\times} e^{-i(\Omega t + k_{\Omega}z)}(\partial_{x} E_{y} + \partial_{y} E_{x}) = 0,
 \nonumber \\
\nu = 1:\quad & \partial_{ct} E_{x} - (\mbox{\boldmath$\nabla$}  \times \mathbf {B})_{x} + h_{+}e^{-i(\Omega t + k_{\Omega}z)} \nonumber\\
& \quad \times (i\Omega E_{x} + ik_{\Omega}B_{y} - \partial_{ct} E_{x} - \partial_{z} B_{y}) \nonumber \\
 &  +  h_{\times}e^{-i(\Omega t + k_{\Omega}z)} (i\Omega E_{y} + ik_{\Omega}B_x - \partial_{ct} E_{y} + \partial_{z} B_{x})\nonumber\\
 &\quad = 0,\nonumber \\
\nu = 2:\quad & \partial_{ct} E_{y} - (\mbox{\boldmath$\nabla$}  \times \mathbf {B})_{y} - h_{+}e^{-i(\Omega t + k_{\Omega}z)}\nonumber\\
&\quad \times (i\Omega E_{y} + ik_{\Omega}B_{x} + \partial_{z} B_{x}-\partial_{ct} E_{y})\nonumber \\
& -  h_{\times}e^{-i(\Omega t + k_{\Omega}z)} (i\Omega E_{x} + ik_{\Omega}B_y - \partial_{ct} E_{x} - \partial_{z} B_{y})\nonumber\\
&\quad = 0,\nonumber \\
\nu = 3:\quad & \partial_{ct} E_{z} - (\mbox{\boldmath$\nabla$}  \times \mathbf {B})_{z} + h_{+}e^{-i(\Omega t + k_{\Omega}z)} (\partial_{x} B_{y} + \partial_{y} B_{x})\nonumber\\
& \quad +  h_{\times}e^{-i(\Omega t + k_{\Omega}z)} (\partial_{x}
B_{x} - \partial_{y} B_{y}) = 0.\label{expandedeqn}
\end{align}

\noindent Since $\Omega \ll \omega_{0}$ we can drop the terms in
$\Omega$ and $k_{\Omega}$. We also make use of the condition $z = 0$
and choose \mbox{$h_{+} \not= 0, h_{\times} = 0$.} This leads to the
simplified equations

\begin{align}
\nu = 0:\quad &\mbox{\boldmath$\nabla$}  \cdot \mathbf {E} = h_{+}e^{-i\Omega t } (\partial_{x} E_{x} - \partial_{y} E_{y}),\nonumber \\
\nu = 1:\quad &\partial_{ct} E_{x} - (\mbox{\boldmath$\nabla$}  \times \mathbf {B})_{x} = h_{+}e^{-i\Omega t } (\partial_{ct} E_{x} + \partial_{z} B_{y}), \nonumber \\
\nu = 2:\quad &\partial_{ct} E_{y} - (\mbox{\boldmath$\nabla$}  \times \mathbf {B})_{y} = h_{+}e^{-i\Omega t } (\partial_{z} B_{x}-\partial_{ct} E_{y}), \nonumber \\
\nu = 3:\quad &\partial_{ct} E_{z} - (\mbox{\boldmath$\nabla$}
\times \mathbf {B})_{z} = - h_{+}e^{-i\Omega t } ( \partial_{x}
B_{y} + \partial_{y} B_{x}).\label{lineareqn}
\end{align}

\noindent In the next section we will show how to solve these equations when a plane (light) wave of angular frequency $\omega_{0}$ propagates along the $x$ and/or $y$ axis.

\section{Solution of the equations of motion}

The electromagnetic (em) fields $\mathbf  E, \mathbf  B$ in the arms consist of the carrier field $\mathbf  E_{0},
 \mathbf  B_{0}$ due to the external source (the laser beam),
and the additional field $\mathbf  E_{+}, \mathbf  B_{+}$, generated by the source terms in Eq. \eqref{lineareqn},

\begin{equation}
\mathbf  E = \mathbf  E_{0} + \mathbf  E_{+}, \qquad  \mathbf  B = \mathbf  B_{0} + \mathbf  B_{+}.
\end{equation}

\noindent We describe the carrier as a plane wave propagating along
the $x$-axis and polarized along the $z$-axis, and of constant amplitude
$A_{0}$

\begin {align}
\mathbf  E_{0} & =  A_{0}e^{-i(\omega_{0}t - k_{0}x)}\mathbf{ u}_{z}, \nonumber\\
\mathbf  B_{0} & = - A_{0}e^{-i(\omega_{0}t - k_{0}x)}\mathbf
{u}_{y}.\label{soln0}
\end{align}

\noindent To the sideband fields $\mathbf  E_{+}, \mathbf  B_{+}$, we
assign slowly time-dependent amplitudes $E_{+}(t), B_{+}(t)$ and a
phase factor $(\omega_{+}t - k_{0}x)$ where $\omega_{+} = \omega_{0}
+ \Omega$, namely,

\begin {align}
\mathbf  E_{+} & =  E_{+}(t)e^{-i(\omega_{+}t - k_{0}x)}\mathbf {u}_{z}, \nonumber\\
\mathbf  B_{+} & = B_{+}(t)e^{-i(\omega_{+}t - k_{0}x)}\mathbf
{u}_{y}.\label{soln+}
\end{align}

We will solve the equations in \eqref{lineareqn} perturbatively
noting that $E_{+}, B_{+}$ are of order $h$ with respect to $E_{0},
B_{0}$. Thus on the left side of Eq. \eqref{lineareqn} we must use
the full fields $E, B$ but on the right side it suffices to retain
only $E_{0}, B_{0}$. With the choice of Eq. \eqref{soln0} for the
carrier field only the $\nu = 3$ equation is nontrivial; the first
three equations are satisfied automatically with the choices made in
Eqs. (19,20).  Writing out the part of order $h$, for the $\nu = 3$
equation in (17), gives

\begin{align}
- \frac{\partial E_{+z}}{c\partial {t}} + \frac{\partial
B_{+y}}{\partial x} & = h_{+}e^{-i\Omega t}\frac{\partial B_{0y}}{\partial x}\nonumber\\
& = -
ih_{+}k_{0}A_{0} e^{-i\omega_{+} t} e^{ik_{0}x}.\label{1st}
\end{align}

\noindent  The remaining terms of the equation
\begin{equation}
-\frac{\partial E_{0z}}{c\partial t} + \frac{\partial B_{0y}}{\partial x} = 0,
\end{equation}

\noindent describe the
propagation of the free carrier field. The fields, both carrier and
sideband satisfy Maxwell's dual equation, namely  $\partial_{\mu}\tilde F^{\mu \nu} = 0$.
For our choice of direction of propagation and polarization, this condition
reduces to

\begin{equation}
- \frac{\partial E_{+z}}{\partial x} + \frac{\partial B_{+y}}{c\partial t} = 0.\label{2nd}
\end{equation}

\noindent Combining Eqs. \eqref{1st} and \eqref{2nd} leads to wave equations for $E_{+z}, B_{+y}$

\begin{eqnarray}
\frac{\partial ^{2} E_{+z}}{\partial x^{2}} - \frac{\partial ^{2}
E_{+z}}{c^{2}\partial t^{2}} = - h_{+} k_{+} k_{0} A_{0} e^{-i(\omega_{+}t - k_{0} x)},\nonumber\\
\frac{\partial ^{2} B_{+y}}{\partial x^{2}} - \frac{\partial ^{2}
B_{+y}}{c^{2}\partial t^{2}} = + h_{+} k_{0}^{2} A_{0} e^{-i(\omega_{+}t - k_{0} x)},
\end{eqnarray}

\noindent where we have introduced $k_{+} = \omega_{+}/c$.

We now use Eqs. \eqref{soln+} and keep in Eqs. (24) only the first
time derivatives  ${dE_{+}(t)}/{dt},  {dB_{+}(t)}/{dt}$ of the
amplitudes, since $E_{+}(t), B_{+}(t)$ vary slowly. We can also
approximate $\omega_{+}^{2}/c^{2} - k_{0}^{2} \simeq 2\Omega
\omega_0/c^{2}$ and find the following two eqs for the slowly
varying amplitudes

\begin{eqnarray}
\frac{dE_{+}}{dt} - i\frac{\omega_0}{\omega_{+}}\Omega E_{+} = i \frac{h_{+}\omega_{0} A_{0}}{2},\nonumber\\
\frac{dB_{+}}{dt} - i\frac{\omega_0}{\omega_{+}}\Omega B_{+} = - i \frac{h_{+}\omega_{0}^{2}A_{0}}
{2\omega_{+}},
\end{eqnarray}

\noindent with solutions

\begin{eqnarray}
E_{+}(t) = \frac{\omega_{+}}{\omega_{0}}\frac{h_{+}\omega_{0} A_{0}}{2 \Omega}
(e^{i\frac{\omega_{0}}{\omega_{+}}\,\Omega t} -1),\nonumber\\
B_{+}(t) = - \frac{h_{+}\omega_{0} A_{0}}{2 \Omega}
(e^{i \frac{\omega_{0}}{\omega_{+}}\,\Omega t} -1).
\end{eqnarray}

\noindent where we assumed the initial conditions $E_{+}(t=0) =
B_{+}(t=0) =0$. Therefore, to the approximation $\Omega \ll \omega_{0}$
that we are using, the sideband fields are

\begin{eqnarray}
\mathbf {E}_{+} = \frac{i\omega_{0}}{\Omega} h_{+} A_{0} e^{\frac{i \Omega
t}{2}} \sin \frac{\Omega t}{2}\, e^{-i(\omega_{+}t -k_{0}x)}\mathbf
{u}_{z},\nonumber\\
\mathbf {B}_{+} = -\frac{i\omega_{0}}{\Omega} h_{+} A_{0} e^{\frac{i \Omega
t}{2}} \sin \frac{\Omega t}{2}\, e^{-i(\omega_{+}t -k_{0}x)}\mathbf
{u}_{y}.
\end{eqnarray}

\noindent The amplitudes $E_{+}(t), B_{+}(t)$ of the sideband fields
are zero at $t=0$ and grow in time as a parametric amplification
process. We will designate them by $A_{+}(t)$. The maximum value of
$A_{+}(t)$ is determined by the losses in the arm cavities, however
the carrier amplitude, $A_{0}$, remains constant. For one round trip
$t=2L/c = 2T$ and the sideband amplitude has the value

\begin{equation}
A_{+}(2T) =i h_{+}L k_{0}  A_{0} e^{i\Omega T}\frac{\sin \Omega
T}{\Omega T}.
\end{equation}

We see that after one round trip the amplitude $A_{0}$ of the carrier
acquires a small complex part

\begin{equation}
A_{+}(2T) = iA_{0} \phi_{A+},
\end{equation}
where
\begin{equation}
\phi_{A+} = \phi_{A}e^{i\Omega T} = h_{+} k_{0}L \frac{\sin \Omega T}
{\Omega T} e^{i\Omega T} \ll 1.
\end{equation}

\noindent Thus the amplitude of the propagating light can be written as

\begin{equation}
A = A_{0} + iA_{0}\phi_{A+} \simeq A_{0}e^{i\phi_{A+}},
\end{equation}

\noindent which shows that the phase of the carrier is shifted
 by exactly the same amount as was calculated
in section 2. When the carrier propagates in the $y$-arm
the sign of $h$ and thus also of $A_{+}(2T)$ is reversed;
after subtracting the fields returning from the two arms,
we find for their relative phase difference  $2\phi_{A+}$, as
we had obtained in Eq. (10).\\

Consider now a gw that depends sinusoidally on time, which we
express at $z = 0$ by

\begin{equation}
h_{+}(t) = h_{+} \cos \Omega t = h_{+}\left(\frac{e^{i\Omega t} +
e^{-i\Omega t}}{2}\right).
\end{equation}

\noindent It is clear that in this case both an upper and lower sideband
will be present. The fields are

\begin{equation}
E_{+} = \frac{iE_{0} \phi_{A+}}{2}\,e^{-i(\omega_{+}t - k_0 x)},
\quad E_{-} = \frac{iE_{0} \phi_{A-}}{2}\,e^{-i(\omega_{-}t - k_0
x)},
\end{equation}

\noindent where $\phi_{A-} = \phi^{*}_{A+} = \phi_{A}e^{-i\Omega
T}$; to see this change \mbox{$\Omega \rightarrow - \Omega$} in
Eq.(30).
 At $t=0, x =0$ the two sideband
fields have equal imaginary parts and opposite real parts.
 This is the condition that corresponds to phase
(or frequency) modulation when the sidebands are combined with the
carrier

\begin{eqnarray}
E & = & E_{0} e^{-i (\omega_{0} t - k_0 x)}\left(1 +
\frac{i\phi_{A+}}{2}e^{-i\Omega t}
 + \frac{i\phi_{A-}}{2}e^{i\Omega t}\right) \nonumber\\
  & = & E_{0} e^{-i (\omega_{0} t - k_0 x)}\left[1 + i\frac{\phi_{A}}{2} (e^{-i\Omega(t - T)}
  + e^{i\Omega(t-T)}) \right]
 \nonumber\\
& \simeq & E_{0} e^{-i[ \omega_{0} t - \phi_{A} \cos \Omega (t - T)
- k_0 x]}.
\end{eqnarray}

\noindent In conclusion a sinusoidal gw interacting with a plane wave carrier
(under the appropriate geometry) imposes upper and lower sidebands,
or equivalently contributes a time-dependent phase shift. We
obtained this result by considering the propagation of the carrier
in the space-time of the gw, without referring to the change in the
round-trip time of travel to and from the end mirror.

\section {Discussion}

Our derivation is based on a {\bf{subtle point}},  the use of the TT
gauge. We can use the TT gauge only because the end points of the
light travel (the mirrors) are free. If the mirrors were fixed we
would have to do the calculation in the LL gauge, i.e. in the
laboratory frame of reference. In the LL gauge
\cite{Thorne,Malik2,Pegoraro,Tarabrin} the amplitude of the gw is
modified from the form given by Eq. (4), the relevant part of the
metric being

\begin{equation}
h_{00} = \frac{1}{2c^2} \ddot{h}(t)(x^2 -y^2).
\end{equation}

For low frequency gw's the resulting phase shift when the mirrors
are fixed is much smaller than for free mirrors, and this is why
interferometric gw detectors are constructed with suspended mirrors.
For the details of
the calculation in the LL gauge see ref. \cite{Malik2}. \\

The presence of sidebands has a direct physical interpretation in
terms of absorption and stimulated emission of gravitons from/to the
gw. Since the gw field is highly classical, [the occupation number
for $h \sim 10^{-23}$ and $f \sim 100$ Hz, is $n = N_g/(\lambda
/2\pi)^{3} \sim 10^{33}$], both processes have the same probability.
Absorption leads to the upper sideband, while emission to the lower
one. There is no energy exchange between the optical and
gravitational fields, but
only a phase shift.\\

In our analysis we have assumed that the carrier is a plane wave
propagating toward the positive x-direction, see Eqs. (19). In
practice the carrier is reflected by a mirror at $x=a$, thus the
Electric field should vanish at $x=a$. This condition is satisfied
by writing the fields in the form

\begin {align}
\mathbf  E_{0} & =  A_{0}[e^{-i(\omega_{0}t - k_{0}x)} - e^{-i(\omega_{0}t
 + k_{0}x - 2 k_{0}a)}]\mathbf{ u}_{z}, \nonumber\\
\mathbf  B_{0} & = - A_{0}[e^{-i(\omega_{0}t - k_{0}x)} +
e^{-i(\omega_{0}t + k_{0}x) - 2 k_{0}a}]\mathbf
{u}_{y}.\label{soln00}
\end{align}

If a mirror is also placed at $x=0$ to reflect the carrier and
$k_{0}a =n\pi$, a standing wave is established in the region $0
\leqq x \leqq a$. This does not modify the conclusions that we have
reached. The case when the carrier is a standing wave is treated by
Cooperstock and Faraoni \cite{Cooperstock2} and leads to the same
results as in Eq. (34). The interaction of an electromagnetic and
gravitational field was first discussed by Gertsenshtein and
Pustovoit \cite{Gertsenshtein} in 1963, and
subsequently by others \cite{Caves, Pegoraro, Cooperstock1, Lobo}.\\

We have also shown that a gw will couple to the carrier in a single
arm. The interferometer configuration has been chosen because it is
technically advantageous. The light returning from the two arms is
adjusted to interfere destructively at the detection point in the
absence of a gw. Thus when the gw induces a phase shift, a signal
appears over a null background (excluding noise). By using multiple
traversals in the arms one can increase the effective length, and
thus the phase shift, significantly. The main limitation in
effective arm length (apart from losses in the optics) is related to
the frequency of the gw. We have seen in Eq. (10) that the phase
shift is modulated by the ``form factor" sin$(\Omega T)/\Omega T$,
and therefore $\Omega L/c$ must be kept small.

 We thank Dr. R. Weiss for bringing to our attention  ref.
 \cite{Cooperstock2}. This work was supported in part by
 DOE Grant DE-FG02-91ER40685 and
NSF Grant PHY-0456239.

\end{document}